%% file: main.tex
\documentclass[11pt]{article}

\input{preamble}

\cbdctrue
\visaconfidentialfalse

\renewcommand{\authnote}[2]{}

\begin{document}
	\sloppy
	
	\ifvisaconfidential
		\pagestyle{myfancy}			
	\fi
	
	\title{Towards a Two-Tier Hierarchical Infrastructure: \\An Offline Payment System for Central Bank Digital Currencies}
	
	
	\author[*]{\small Mihai Christodorescu}
	\author[**]{Wanyun Catherine Gu}
	\author[*]{Ranjit Kumaresan}
	\author[*]{Mohsen Minaei}
	\author[*]{Mustafa Ozdayi}
	\author[**]{Benjamin Price}
	\author[*]{Srinivasan Raghuraman}
	\author[*]{Muhammad Saad}
	\author[**]{Cuy Sheffield}
	\author[*]{Minghua Xu}
	\author[*]{Mahdi Zamani}
	\affil[*]{Visa Research, Palo Alto, CA}
	\affil[**]{Visa Crypto Product, Palo Alto, CA}
	
	\date{}
	\maketitle
	
	\ifvisaconfidential
		\thispagestyle{myfancy}		
	\fi
	
	\input{abstract}

	\newpage
	\begin{doublespacing}
		\tableofcontents
	\end{doublespacing}
	
	\newpage
	\input{intro}
	\input{problem}
	\input{basic-protocol}

	\bibliographystyle{unsrt}
	\bibliography{main}
	
	\input{disclaimers}
\end{document}

%% file: preamble.tex
\usepackage[tt=false, type1=true]{libertine}
\usepackage[T1]{fontenc}
\usepackage[utf8]{inputenc}
\usepackage[libertine]{newtxmath}

\usepackage{import}
\usepackage{url}
\usepackage[margin=1in]{geometry}
\usepackage{xcolor}
\usepackage{xspace}
\usepackage{comment}
\usepackage{setspace}

\definecolor{darkblue}{rgb}{0.0,0.0,0.75}
\definecolor{lightgray}{rgb}{0.95,0.95,0.95}
\usepackage[bookmarks=true,bookmarksnumbered=false,bookmarksopen=true,hidelinks=false,colorlinks=true,citecolor=darkblue,linkcolor=darkblue]{hyperref}

\usepackage{paralist}
\usepackage{rotating}
\usepackage{graphicx}
\usepackage{bookmark}
\usepackage[sort]{cite}
\usepackage{epstopdf}
\usepackage{mathtools}
\usepackage{enumitem}
\usepackage{algorithmic}
\usepackage[operators, sets, adversary] {cryptocode}
\usepackage{framed}
\usepackage{float}
\usepackage{booktabs}
\usepackage{color,colortbl}
\usepackage{amsfonts}
\usepackage{relsize}
\usepackage{subfigure}
\usepackage[affil-it]{authblk}
\usepackage[capitalise, noabbrev]{cleveref}

\setlist[description]{listparindent=\parindent,leftmargin=0em,itemsep=0.3em,topsep=0.3em}

\setbox1=\hbox{$\bullet$}\setbox2=\hbox{\tiny$\bullet$}

\usepackage[framemethod=tikz]{mdframed}
\usetikzlibrary{shadows}
\newmdenv[
	roundcorner=2pt,
	outerlinewidth=0.25pt,
	linecolor=black,
	shadow=true,
	shadowsize=4pt,
	frametitlerule=true,
	frametitlerulewidth=0.25pt,
	frametitlebackgroundcolor=gray!17,
	frametitlealignment={\vspace{0.003\linewidth}\center},
	font=\fontsize{9.75}{11.5}\selectfont,
	nobreak=true,
]{shadowbox}

\newcommand{\formatNode}[1]		{\ensuremath{\mathsf{#1}}\xspace}
\newcommand{\A}					{\formatNode{A}}		
\newcommand{\B}					{\formatNode{B}}		
\newcommand{\C}					{\formatNode{C}}		
\renewcommand{\S}				{\formatNode{S}}		
\newcommand{\M}					{\formatNode{M}}		
\newcommand{\PS}				{\formatNode{PS}}		
\newcommand{\OPS}				{\formatNode{OPS}}		
		
\newcommand{\D}					{\formatNode{D}}

\newcommand{\SH}				{\ensuremath{\mathsf{D}}\xspace}	
\newcommand{\SE}				{\ensuremath{\mathsf{T}}\xspace}	

\newcommand{\formatTag}[1]		{\ensuremath{\mathrm{#1}}}
\newcommand{\sender}			{\formatTag{sender}}

\newcommand{\formatVar}[1]		{\ensuremath{\mathsf{#1}}\xspace}

\newcommand{\bal}				{\formatVar{bal}}

\newcommand{\onbal}				{\formatVar{onBal}}
\newcommand{\offbal}			{\formatVar{offBal}}
\newcommand{\sk}				{\formatVar{sk}}
\newcommand{\vk}				{\formatVar{vk}}
\newcommand{\sig}				{\formatVar{sig}}
\newcommand{\cert}				{\formatVar{cert}}
\renewcommand{\index}			{\formatVar{index}}
\newcommand{\id}				{\formatVar{i}}
\newcommand{\pid}				{\formatVar{j}}
\newcommand{\userregistry}		{\formatVar{Registry}}				

\newcommand{\iplog}				{\formatVar{inPaymentLog}}			
\newcommand{\plog}				{\formatVar{paymentLog}}			

\newcommand{\amount}			{\formatVar{amount}}
\newcommand{\type}				{\formatVar{type}}
\newcommand{\ptime}				{\formatVar{time}}

\newcommand{\receiver}			{\formatVar{receiver}}

\newcommand{\formatFunc}[1]		{\ensuremath{\mathtt{#1}}\xspace}
\renewcommand{\H}				{\formatFunc{H}}
\newcommand{\payVerify}			{\mathtt{PayVerify}}

\newcommand{\keygen}			{\mathsf{KeyGen}}
\newcommand{\sign}				{\mathtt{Sign}}
\newcommand{\sigverify}			{\mathtt{SigVerify}}
\newcommand{\certinit}			{\mathtt{CertInit}}
\newcommand{\certverify}		{\mathtt{CertVerify}}
\newcommand{\hwcertverify}		{\mathtt{TACertVerify}}
\newcommand{\oemcertverify}		{\mathtt{OEMVerify}}

\newcommand{\init}				{\formatFunc{Init}}

\newcommand{\deposit}			{\formatFunc{Deposit}}
\newcommand{\pay}				{\formatFunc{Pay}}

\newcommand{\collect}			{\formatFunc{Collect}}

\newcommand{\withdraw}			{\formatFunc{Withdraw}}

\newcommand{\getbalance}		{\formatFunc{Get\text{-}Balance}}

\newcommand{\formatMsg}[1]		{\ensuremath{\mathbf{#1}}\xspace}
\newcommand{\clientRegisterMsg}	{\formatMsg{RegisterClient}}
\newcommand{\taRegisterMsg}		{\formatMsg{TARegister}}
\newcommand{\depositMsg}		{\formatMsg{Deposit}}
\newcommand{\depositConfirmed}	{\formatMsg{DepositConfirmed}}
\newcommand{\payMsg}			{\formatMsg{RequestPayment}}
\newcommand{\payConfirmed}		{\formatMsg{ReceivedPayment}}
\newcommand{\withdrawMsg}		{\formatMsg{Withdraw}}
\newcommand{\withdrawConfirmed}	{\formatMsg{WithdrawConfirmed}}
\newcommand{\claimMsg}			{\formatMsg{Claim}}

\newcommand{\claimConfirmed}	{\formatMsg{ClaimConfirmed}}

\newcommand{\formatStr}[1]		{\text{``}\ensuremath{\mathtt{#1}}\text{''}\xspace}
\newcommand{\basic}				{\formatStr{Basic}}
\newcommand{\conditional}		{\formatStr{Conditional}}
\newcommand{\TA}				{\formatStr{TA}}
\newcommand{\UA}				{\formatStr{UA}}

\newcommand{\authnote}[2]		{\textcolor{blue}{[#1: #2]}}

\newcommand{\mahdi}[1]			{\authnote{Mahdi}{#1}}
\newcommand{\srini}[1]			{\authnote{Srini}{#1}}
\newcommand{\ranjit}[1]			{\authnote{Ranjit}{#1}}
\newcommand{\mohsen}[1]			{\authnote{Mohsen}{#1}}

\newcommand{\orr}			{\thinspace or \thinspace}
\newcommand{\verify}		{\stackrel{?}{=}}

\newcommand{\ignore}[1]		{}

\usepackage{fancyhdr}
\fancypagestyle{myfancy}{
	\lhead{}
	\rhead{}
	\chead{\textit{Visa Confidential}}
}
\newif\ifcbdc
\newif\ifvisaconfidential

%% file: abstract.tex
\begin{abstract}
	Digital payments traditionally rely on online communications with several intermediaries such as banks, payment networks, and payment processors in order to authorize and process payment transactions. While these communication networks are designed to be highly available with continuous uptime, there may be times when an end-user experiences little or no access to network connectivity.
	
	The growing interest in digital forms of payments has led central banks around the world to explore the possibility of issuing a new type of central-bank money, known as \emph{central bank digital currency~(CBDC)}. To facilitate the secure issuance and transfer of CBDC, we envision a CBDC design under a two-tier \textit{hierarchical trust infrastructure}, which is implemented using public-key cryptography with the central bank as the root certificate authority for generating digital signatures, and other financial institutions as intermediate certificate authorities. One important design feature for CBDC that can be developed under this hierarchical trust infrastructure is an ``offline'' capability to create secure point-to-point offline payments through the use of authorized hardware. An offline capability for CBDC as digital cash can create a resilient payment system for consumers and businesses to transact in any situation. 
	
	In this paper, we propose an \emph{offline payment system (OPS)} protocol for CBDC that allows a user to make digital payments to another user while both users are temporarily offline and unable to connect to payment intermediaries (or even the Internet). OPS can be used to \emph{instantly} complete a transaction involving any form of digital currency over a point-to-point channel without communicating with any payment intermediary, achieving virtually unbounded throughput and real-time transaction latency. One needs to ensure funds cannot be double-spent during offline payments as no trusted intermediary is present in the payment loop to protect against replay of payment transactions. Our OPS protocol prevents double-spending by relying on digital signatures generated by \emph{trusted execution environments (TEEs)} which are already available on most computer devices, including smartphones and tablets. While a TEE brings the primary point of trust to an offline device, an OPS system requires several cryptographic protocols to enable the secure exchange of funds between multiple TEE-enabled devices, and hence a reliable financial ecosystem that can securely support CBDC at scale.

\end{abstract}

%% file: intro.tex
\section{Introduction} \label{sec:intro}
Digital payments today represent an account-based system of debiting and crediting accounts operated by the financial institutions, where the ownership of a payment account is tied to a user's public identity. With the emergence of distributed ledger technology, there has been growing interest in a new form of token-based digital payment, where the token itself represents the medium of exchange or money, and the ownership of the token is determined by a user's access to a private cryptographic key that provides access to the user's \textit{digital wallet}. Access to these wallets are typically facilitated by entities, known as \textit{wallet providers}, that offer secure access to the cryptographic keys as well as some banking and other financial capabilities~\cite{uscongress:online,Cryptocu50:online,Investig5:online}.

The growing interest in token-based payments have led central banks around the world to explore the possibility of issuing a new type of central-bank money, known as \emph{central bank digital currency~(CBDC)}. Some of these central banks have proposed designs that would issue this new money in the form of cryptographic tokens backed directly by central bank reserves to enable consumers and businesses to make payments in the form of ``digital cash''~\cite{BoE:2020,Riksbank:2020,PBoC:2020}.

In a CBDC model, the money in transit should remain the liability of its trusted issuer (e.g., a central bank in the case of CBDC), meaning that its value is always guaranteed by the issuer as long as the recipient can easily verify the authenticity of the money. This ensures that (1) the money was issued properly according to a specific set of rules (aka, a monetary policy that has parallels to the existing financial system) set by the issuer, and (2) the money maintained its value in transit (i.e., was neither double spent nor counterfeited). 
In the digital world, this can be implemented using public-key cryptography, where the money in transit carries a digital signature that can only be generated directly by the central bank or indirectly by one of the central bank's certified delegates. Any recipient of the money can then simply authenticate it by verifying the signature against the public key of the central bank and/or the certificate of its trusted delegate. Public-key cryptography could offer significant advantages for security and compliance over cash.

We envision a two-tier \textit{hierarchical trust infrastructure} delivered through certified delegation, which allows the central bank to outsource the complexity of managing digital certificates for CBDC tokens to a set of potentially regulated, permissioned entities that derive their authority from the central bank, through a hierarchy of digital certificates originated from the central back at the root. This hierarchical trust design resembles the hierarchy of \emph{certificate authorities (CAs)} in a \emph{public-key infrastructure (PKI)}~\cite{rfc3647} that plays a vital role in enabling the secure transfer of information over the Internet. 
A PKI model for CBDC can significantly facilitate the secure issuance and transfer of CBDC funds with the central bank serving as the root CA and supervised financial institutions (FIs) serving as intermediate CAs under regulatory oversight. These intermediate CAs have two roles: (1) Vetting wallet providers based on regulatory compliance; (2) Issuing digital certificates to vetted wallet providers to facilitate CBDC payments securely.

The core advantage of the two-tier model is that it decouples the certificate infrastructure (Tier 1) from the critical latency path of CBDC payments (Tier 2), allowing wallet providers such as banks and other FIs to securely process CBDC payments at a high scale without imposing extra overhead on the highly-protected PKI nodes. This is a particularly relevant question at the heart of the current CBDC debate. 

In the past year, several central banks have begun to research and ask how offline CBDC payments could occur~\cite{TheFedCo13:online,BoE:2020,Thetechn45:online,jp70:online}.	
\textbf{An important feature that can be developed under the hierarchical trust infrastructure for CBDC is an “offline” capability to create secure point-to-point offline payments using authorized hardware.} An offline protocol can be a potential feature of CBDC by bringing the primary point of trust to an offline device under this infrastructure. CBDC as digital cash can move instantly across multiple payment rails and condition without necessarily needing to directly involve any intermediary during transfers. For example, if the sender and the recipient of a payment have relationships with different wallet providers, they should still be able to transact with each other instantly in an ad-hoc, point-to-point fashion without communicating with their wallet providers. This enables a significantly higher throughput as payments can still happen under congested network conditions. Moreover, the design provides a higher level of privacy to the clients by avoiding the need to share unnecessary payment information with the intermediaries.



One technical challenge in creating secure offline payments is to protect the system from financial crimes and to avoid exposing either the buyer, seller, or the central bank to the risk that the payment may not ultimately be settled~\cite{BoE:2020}.  
Existing card payment networks such as Visa and Mastercard can provide some form of offline payments for situations where the acceptance device (e.g., a card terminal) cannot connect to payment providers for authorization in real time~\cite{visa:merchant:guide:2015,mastercard:guide:2019}.
%
%
%
Payments without issuer authorization require the merchant to bear some counterparty risk because the payer may not actually have the necessary funds to fulfill the transaction~\cite{square:offline}.
%
%
\begin{quote}
	\emph{Would it be possible to allow offline payments between two parties without exposing counterparty risk on either party?}
\end{quote}

Today, most mobile devices (e.g., smartphones and tablets) are equipped with \emph{secure hardware} to store keys and other sensitive material that can only be accessed through strong user authentication measures (e.g., biometrics). It has been shown that compromising these hardware-protected mobile devices without help from their manufacturers is difficult~\cite{secure:element:difficulty}. This can potentially make mobile devices a viable option to store the user's CBDC funds and to issue offline payments using hardware-protected credentials provisioned by the central bank or one of its delegates. As long as the secure hardware remains secure, (1) the keys for signing offline payments are protected from malicious access, and (2) the same funds cannot be spent offline more than once (i.e., no double spending).

While secure hardware provides a simple and efficient medium for delegation of trust in the digital setting [14], the possibility of device compromise would not only expose the involved users to the risk of funds loss but also could, at a much larger scale, jeopardize the functionality of the entire CBDC ecosystem. In a recent study, Allen~et~al.~\cite{cbdc:brookings} identify three main challenges in the use of secure hardware. First, there is a strong economic incentive for users to compromise their secure hardware in order to counterfeit CBDC funds. Second, compromising one device could allow a user to double spend funds an unlimited number of times (i.e., by default there is no \emph{graceful degradation}). And third, a user’s funds could be totally lost in case of device loss (due to, e.g., damage or failure).

We observe that these challenges are not exclusive to digital offline payments and are applicable to virtually any form of offline payments, including physical cash. Therefore, we envision that measures similar to those used for physical cash could be employed to protect the security of CBDC in an offline setting while still maintaining the practical benefits of offline payments. In this paper, we initiate the formal study of offline digital payments and propose a system that provides the basic functionalities for offline payments assuming secure hardware cannot be compromised. In subsequent work, we will explore extended techniques for 
offline payments to alleviate the above challenges by addressing economic incentives, graceful degradation, and funds recovery.

\subsection{Our Contribution}
In this paper, we initiate the study of offline digital payments by defining the notion of an \emph{offline payment system (OPS)} that allows a user (e.g., a customer) to make digital payments in CBDCs to another user (e.g., a merchant) while both users are temporarily offline from payment intermediaries (or even the Internet). We then construct the first OPS protocol that allows point-to-point authorization of offline payments using open source technology and public key infrastructure to significantly reduce the overhead of onboarding new participants compared to existing digital payment systems. Once provisioned, OPS wallets can securely sign and transmit transaction messages directly with each other over any communication channel they prefer without requiring an intermediary to authorize and settle it. Recipients can submit signed, offline payment messages to an authorized wallet provider with guaranteed settlement of those transactions in order to withdraw funds from the offline payment system. 


\input{overview}

%% file: overview.tex
\subsection{Overview of Our Solution}


Consider two \emph{clients} $\A$ and $\B$ who hold online accounts with a \emph{server} $\S$. We assume an account maintains information about the amount of money (aka, balance) that the client holds at $\S$.
\ifcbdc
	We assume that the server is a digital wallet provider that has already obtained a digital certificate from the central bank through the hierarchical trust infrastructure for CBDC, as described in Section~\ref{sec:intro}. This certificate can be used to attest to the clients that $\S$ is a trusted entity. In the rest of this paper, we describe the offline payment protocol that happens between the clients and the trusted server.
\fi

We assume a client's account maintains information about the amount of money (aka, balance) that the client holds at $\S$.
The goal of our OPS protocol is to allow client $\A$ (aka, the \emph{sender}) to pay client $\B$ (aka, the \emph{receiver}) an amount of money denoted by $x$ from $\A$'s account with $\S$ without either client communicating with $\S$ during the payment.
We assume $\A$ (or any other client who wishes to send money) owns a secure device $\SH_\A$ that can securely store data and execute code via a \emph{trusted execution environment (TEE)}. However, we do not require $\B$ (or any other client who only wants to receive money) to own a secure device. 
In the following, we first describe our TEE model and the main components of our protocol. Next, we briefly describe the main OPS protocols to set up the clients and perform offline payments.

\begin{description}
	\item[TEE Model.] 
	A TEE is a software stack stored on a read-only memory within a secure device. This software stack should adhere to privacy, security, and high software development standards which are already adopted by most secure devices today. The stack consists of a set of resources to access the secure device, a \emph{trusted operating system (TOS)} that provides developer access to the underlying secure device, and one or more \emph{trusted applications (TA)} that implement application-specific functionalities to be executed securely by the TEE (see Figure~\ref{fig:opsTA}).
	
	\item[Untrusted Applications.] Every client (with or without TEE) has an application-specific \emph{untrusted application~(UA)} that resides in the ``untrusted'' region of the device and thus could be malicious. A benign UA provides user-facing functionalities to receive, verify, and store payments on the device as well as to submit the payments to the server whenever the client goes online. In case of a TEE-enabled client, the UA also interacts with the TA to provide wallet operations to the user, such as creating new offline payments, adding/collecting received payments into the secure wallet, etc. (see Section~\ref{sec:teemodel} for more details).
	
	\item[OPS Components.] Our OPS consists of the following four main components:
	\begin{itemize}[itemsep=0.5em]
		\item  \textit{OPS Server TA:} Deployed on the server and provides the functionalities to register and set up client devices and manage client accounts.
		
		\item \textit{OPS Sender UA:} Deployed on sender's device and provides the OPS user interface to create offline payments by interacting with the OPS TA, and to interact with the server to register the UA and the TEE.
		
		\item \textit{OPS Receiver UA:} Deployed on the receiver's device and provides the OPS user interface to receive and verify offline payments, and to interact with $\S$ to register the UA and claim offline payments. This UA does not interact with any TEE. If the receiver is also wishing to receive money from another client, then she needs to deploy the OPS Receiver UA on her device. 
		
		\item \textit{OPS TA:} Deployed on the sender's secure device (within TEE) and provides OPS-specific functionalities to securely access the secure device and manage the client's offline balance. We denote the TA deployed on client A's device by $\SE_\A$. 
	\end{itemize}
	
	\item[Setup Protocol.] Our OPS protocol requires both clients to register with $\S$ during a one-time, online setup to establish asymmetric cryptographic keys that are later used to issue and verify offline payments. The online setup also allows $\A$ to initialize her TEE jointly by $\S$ and $\SH_\A$'s manufacturer. The TEE setup consists of three phases: (1) Remote attestation to allow either the manufacturer or $\S$ to remotely verify the validity of the TEE stack; (2) TA provisioning to allow either the manufacturer or $\S$ to securely deploy a TA inside TEE; and (3) TA registration to allow the TA to establish a signing key pair, register it with $\S$, and obtain a certificate attesting to the validity of the key pair. See Section~\ref{sec:setup} for the complete description of the setup protocol.
	
	\item[Deposit Protocol.] Client $\A$ needs to initially deposit funds into her secure device when she is online to be able to send offline payments later. Namely, $\A$ requests server $\S$ to deposit an amount of $x$ money from her online balance stored at $\S$ into her offline balance stored in $\SE_\A$. The server responds with a signature on showing that $x$ was deducted from $\A$'s online balance. The client TA verifies the signature with the server's public verification key and adds $x$ to the offline balance stored in $\SE_\A$. See Section~\ref{sec:depositwithdraw} for details.
	
	\item[Offline Payment Protocol.] An offline payment is initiated by the receiver $\B$ who sends a payment request to $\A$, including $\B$'s certificate in the request. Upon receiving the request, $\A$ invokes $\SE_\A.\pay$ to securely deduct the payment amount from $\SE_\A$'s balance and create a signed payment message $P$ containing the payment amount and the certificates of both clients. $\A$ sends $P$ to $\B$ who verifies $\A$'s signature and her certificate, and checks that the payment contains $\B$'s certificate as the recipient. If all checks pass, $\B$ accepts the payment and stores $P$ on his device. Note that by deducting the payment amount from $\A$'s balance (which is stored on the TEE storage), the TEE prevents double spending of that amount. See Section~\ref{sec:offlinepay} for details.
	
	\item[Claim Protocol.] If $\B$ wants to add the amount of $P$ that he received offline from $\A$ to his online balance stored at $\S$, he can invoke the Claim protocol in which $\S$ verifies the validity of $P$ and checks if it was not previously marked as spent using a payment log stored by $\S$. If all checks pass, $\S$ adds the amount of $P$ to $\B$'s online balance and adds $P$ to the log. See Section~\ref{sec:claimcollect} for details.
	
	\item[Collect Protocol.] Imagine that $\B$ also has a secure device with $\SE_\B$ set up similar to what was described before. If $\B$ wishes to make an offline payment out of the money he previously received in $P$ from $\A$ without going online, then he can invoke the Collect protocol to add the money in $P$ into $\SE_\B$'s balance. This allows $\B$ to spend the funds offline in exactly the same way $\A$ made the offline payment $P$. See Section~\ref{sec:claimcollect} for details.
	
	\item[Withdraw Protocol.] If $\A$ wishes to move funds from $\SE_\A$ to her online balance stored at $\S$, then she can invoke the Withdraw protocol which invokes $\SE_\A.\withdraw$ to deduct the funds from $\SE_\A$ and return a message signed with $\SE_\A$'s signing key. The client then forwards the signed withdraw message to $\S$ who adds the fund to $\A$'s online balance after verifying the signature. See Section~\ref{sec:depositwithdraw} for details.
	
	\item[Replay/Rollback Protection.] To protect against malicious intermediaries (such as a malicious UA) replay the messages exchanged between $\S$ and $\SE_\A$ as well as between $\A$ and $\B$, each party maintains monotonically-increasing counters that are incremented after every round of communication between a pair of parties. Both $\S$ and $\SE_\A$ (as well as $\A$ and $\B$) include the latest value of their counter in their signed messages so that the receiver can verify the uniqueness and ordering of all messages according to their local counter which is synchronized after every exchange. 
	
	\mahdi{Add overview of balance recovery and conditional OPS.}
\end{description}

%% file: problem.tex
\section{Our Model}
Consider a group of clients $\A,\B,\C,...$ who can communicate with each other by exchanging messages through a communication network. We assume that a secure communication infrastructure is in place, that is, all parties may interact and send messages to each other in a secure way. In particular, this means that when a party communicates with another, the receiver of the communication will be able to ascertain the authenticity and validity of said communication.

Every client, say $\A$, is associated with a non-negative numeric value known as its \emph{wallet balance} (or simply \emph{balance}) $\bal_\A$ indicating the amount of money possessed by $\A$. A payment is represented in the form ${P: \A \xrightarrow{x} \B}$ indicating a transfer of $x$ amount of money from client $\A$ (aka, the sender) to client $\B$ (aka, the receiver).
A \emph{payment protocol} is a protocol that processes a payment ${P: \A \xrightarrow{x} \B}$ by updating the clients' balances correspondingly, i.e., $\bal_\A = \bal_\A - x$ and $\bal_\B = \bal_\B + x$. A payment is called \emph{authentic} if and only if the sender's balance at the time of payment is greater than or equal to $x$.
A \emph{payment system} $\PS$ consists of a network of clients who can verify the authenticity of any payment within $\PS$ through a designated authority, referred to as \emph{server} $\S$.

\subsection{Problem Definition}
An \emph{offline payment system} (denoted by $\OPS$) is a payment system that enables any pair of clients to pay each other while both are offline from $\S$. More precisely, any client $\A$ is associated with an \emph{online balance} $\onbal_\A$ and an \emph{offline balance} $\offbal_\A$.
Given a payment ${P: \A \xrightarrow{x} \B}$, an \emph{online payment} is a protocol that ensures $\onbal_\A = \onbal_\A - x$ and $\onbal_\B = \onbal_\B + x$.
Given a payment ${P: \A \xrightarrow{x} \B}$, an \emph{offline payment} is a protocol that ensures $\offbal_\A = \offbal_\A - x$ and $\offbal_\B = \offbal_\B + x$.
The following properties are enabled by $\OPS$: \ranjit{OPS does not guarantee offline transitivity in case when client does not have secure element.}

\begin{itemize}
    \item \textbf{Offline Verifiability:} The receiver must be able to independently verify the authenticity of any payment without communicating with the server during the payment.
    
    \item \textbf{Absolute Finality:} Once a payment is complete, the receiver must be instantly guaranteed to own the transferred funds.

    \item \textbf{Online Redeemability:} A client $\A$ must be able to convert any amount $y \leq \offbal_\A$ from their offline balance into their online balance, i.e., $\offbal_\A = \offbal_\A - y$ and $\onbal_\A = \onbal_\A + y$ are executed atomically, and vice versa.     
	
	\item \textbf{Offline Transitivity:} After receiving an offline payment, the receiver must be able to spend the payment amount (or a portion of it) in the same offline session, i.e., without requiring to go online to redeem the payment amount and then spend it.
	
	\item \textbf{Security:} $\OPS$ is secure if it has the following properties:
	\begin{itemize}
		\item \textit{No Double Spending:} No malicious client (or a coalition of them) can spend the same money more than once.
		
		\item \textit{Wallet Security:} No malicious client (or a coalition of them) can spend/remove money from an honest client's wallet without her permission. 
		
		\item \textit{Supply Conservation:} The total supply of money in the system always stays the same, i.e., a client can only add/remove money to/from the system via the deposit/withdraw functionalities provided by the server.
	\end{itemize}	
\end{itemize}

If client $\A$ wishes to spend her money offline, then she must have a TEE-enabled \emph{secure device} $\SH_\A$ to store her offline balance $\offbal_\A$ securely. The balance can only be modified by the \OPS trusted application $\SE_\A$ stored in $\SH_\A$. The authenticity of such modifications are enabled via digital signatures generated by $\SE_\A$ using a secret key $\sk_\A$ stored securely inside $\SH_\A$. If the client does not want to spend money offline, then she does not require any secure device. 

\subsection{Threat Model} 
We assume all parties communicate with each other via secure and authenticated communication channels. We consider a probabilistic, polynomial-time adversary who can corrupt any client in order to (1) prevent the protocol from achieving its defined properties; and (2) counterfeit money, for example, by double-spending the client's money or forging new money.
A corrupt client may do so by arbitrarily tampering with and/or blocking messages exchanged between the server and the client's TEE. We assume that uncorrupt clients employ standard authentication mechanisms such as password and biometrics to prevent unauthorized access to their device in order to spend and/or erase their money without their approval. We finally assume that the server is fully trusted.
\mohsen{we assume all clients are rational, and work to maximize their profits. We allow collusion between clients with the goal of maximizing their utility.}

\input{teeModel}

%% file: teeModel.tex
\subsection{TEE Model} \label{sec:teemodel}
A TEE is an isolated execution environment with its own protected hardware resources (e.g., processor memory, and peripherals) as well as a software stack consisting of an operating system and trusted programs, known as \emph{trusted applications (TAs)}, to access the TEE hardware resources securely~\cite{sabt2015tee,arfaoui2014hood}. The isolation provides strong integrity and confidentiality guarantees, where integrity ensures that unauthorized users cannot change the code of a TA or its behavior, while confidentiality guarantees that  unauthorized access to private TA data is prohibited. The trusted OS provides API access for external programs, known as \emph{untrusted applications (UAs)}, to call and execute public TA functions within the TEE while restricting external access to the rest of the TEE. 

\begin{figure}[t]
    \centering
    \includegraphics[width=0.65\textwidth]{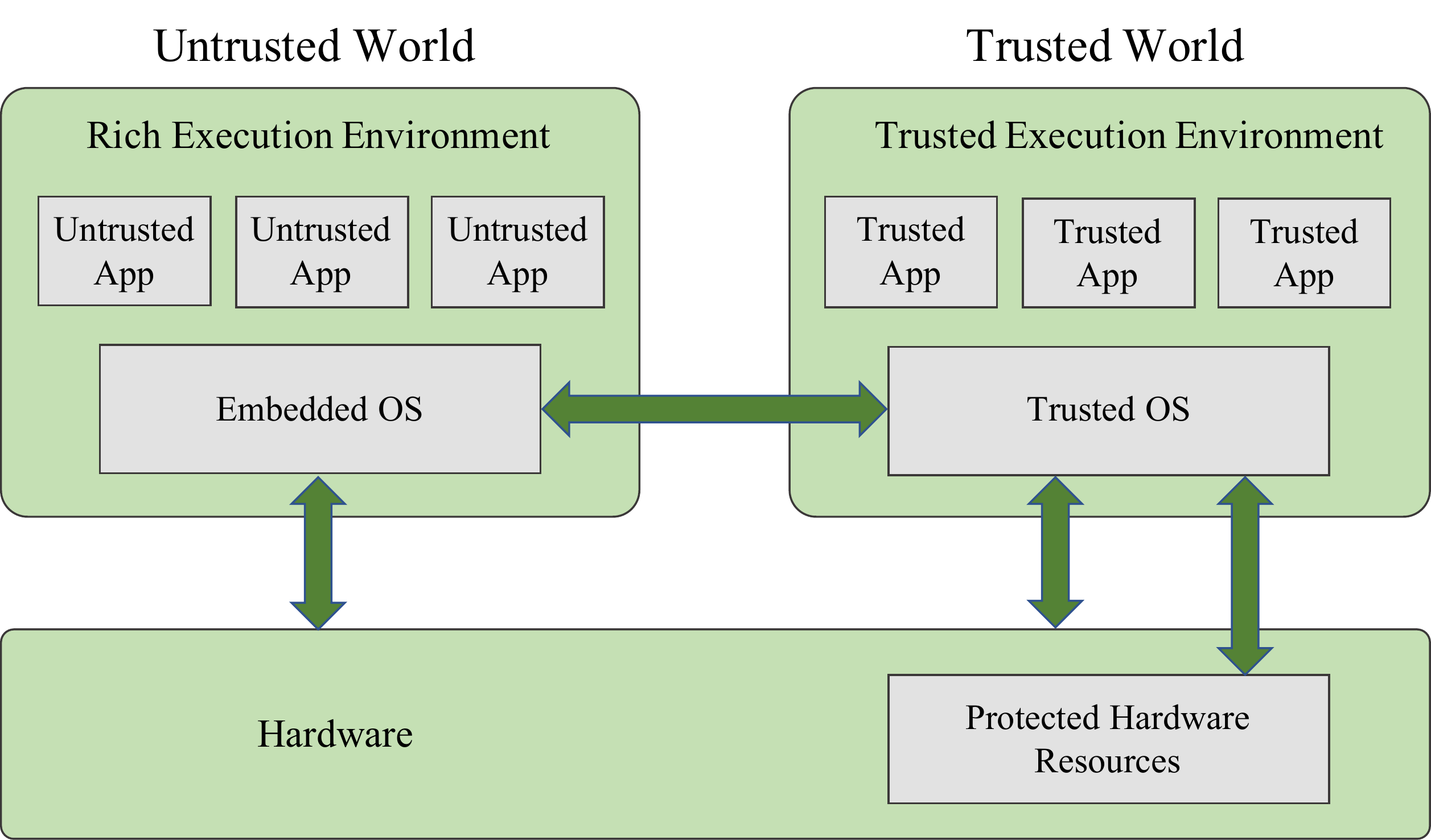}
	\caption{Our TEE Model \srini{to specialize this diagram for OPS? in particular which keys go where.} \mahdi{Specify $\SE_\A$ and $\sk_\A$ as well as secure storage chip in the picture. $\sk_\A$ should reside inside the chip.}}
    \label{fig:teeOverview}
\end{figure}

In practice, there are different TEE architectures depending on the platform they run and the hardware that provides the isolation. In this paper, we target offline payments for mobile devices; therefore, we adopt GlobalPlatform (GP)~\cite{globalWhitePaper}, a standardized TEE model adopted by ARM TrustZone technology~\cite{trustzone,pinto2019trustzone} which itself is used in most Android smartphones today. As shown in~\cref{fig:teeOverview}, the GP model provides a standardized API for UAs in the non-secure world (aka, the \emph{rich execution environment -- REE}) to interact with the isolated TAs via the trusted OS. We now describe GP's secure storage model that allows us to protect against OPS TA state rollback.

\paragraph{TEE Replay/Rollback Protection.} Throughput this paper, we assume that the OPS TA has access to a secure storage to store its state, ensuring that the state cannot be rolled back by the adversary. The GP specification mandates the possibility to store general-purpose data and key material within a TEE with integrity, confidentiality, and atomicity\footnote{~Atomicity means that either the entire write operation completes successfully or no write is done.} guarantees~\cite{GPInternalAPI17}.
 Typically, a \emph{replay-protected memory block (RPMB)} partition\footnote{~The RPMB partition is typically 4 MB in size~\cite{wd:emmc}.} on an eMMC storage (e.g., the phone's persistent storage) is used to store TA's data securely~\cite{wd:emmc,qualcomm}. 
 Any data written on the RPMB is protected against man-in-the-middle replay/rollback attacks using a \emph{monotonically-increasing counter (MIC)} maintained by a dedicated hardware, known as the \emph{RPMB engine}. The engine increments the MIC after every write to the RPMB and uses message authentication codes (MACs) to verify the validity of the write command by checking that (1) the counter was increased, and (2) the MAC that was sent by the sender (e.g., the TA) is identical to the MAC that the RPMB engine generated using its latest value of MIC. Finally, every read from RPMB is MAC-checked by the reader (e.g., the TA) using the latest value of MIC maintained by the reader. For more details, we refer the reader to~\cite{wd:emmc}. 
 The GP specification also allows a secure storage to be implemented on the REE (i.e., non-secure) file system as long as suitable cryptographic protection is applied, which must be as strong as the means used to protect the TEE code and data itself~\cite{GPInternalAPI17}.

%% file: basic-protocol.tex
\newcolumntype{g}{>{\columncolor{lightgray}}p}
\begin{table*}
	\centering
	\renewcommand{\arraystretch}{1.4}
	\setlength{\aboverulesep}{0pt}
	\setlength{\belowrulesep}{0pt}

	\footnotesize
	\begin{tabular}{p{12em} p{30em} p{5em} p{1em}}
		\toprule
		\rowcolor{lightgray}

		\textbf{Variable} &
		\textbf{Description} &
		\textbf{Scope}
		\\ \hline
		
		$\S.\userregistry$		& Server $\S$'s registry of valid UA certificates & $\S$ \\
		$\S.\onbal_\A$			& Client $\A$'s online balance stored at $\S$ & $\S$ \\
		$\S.\id_\A$			& Client $\A$'s index maintained by $\S$ & $\S$ \\ 
		$\SH_\A$			& Client $\A$'s secure device/hardware	& $\A$ \\
		$\SE_\A$			& Client $\A$'s trusted application deployed on $\SH_\A$	& $\A$ \\
		$(\sk_\S, \vk_\S)$  & Server $\S$'s signing key pair 	& ($\S$, Global) \\
		$(\sk_\A, \vk_\A)$  & Client $\A$'s signing key pair 	& ($\A$, Global) \\
		$(\SE_\A.\sk, \SE_\A.\vk)$ & $\SE_\A$'s signing key pair  & ($\SE_\A$, Global) \\
		$\cert_\A$			& Certificate for client $\A$ consisting of $\vk_\A$ and a signature on it by $\S$ certifying that $\vk_\A$ was issued by $\S$ & Global \\
		$\SE_\A.\cert$		& Certificate for $\SE_\A$ consisting of $\SE_\A.\vk$ and a signature on it by $\S$ certifying that $\SE_\A.\vk$ was issued by $\S$ & Global \\
		$\SE_\A.\id$     & Client $\A$'s server index maintained by $\SE_\A$   &  $\SE_\A$\\
        $\SE_\A.\pid$  	 & Client $\A$'s payment index maintained by $\SE_\A$   &  $\SE_\A$ \\		
        $\SE_\A.\bal$    & Client's $\A$'s offline balance & $\SE_\A$ \\ 
        $\SE_\A.\iplog$	 & List of offline payments received by $\SE_\A$ & $\A, \SE_\A, \S$ \\ 
        $P.\amount$		& Amount money transferred by payment $P$ & Holder of $P$ \\
        $P.\sender$	& Certificate of the sender of payment $P$ & Holder of $P$ \\
        $P.\receiver$	& Certificate of the receiver of payment $P$ & Holder of $P$ \\
        $P.\index$		& Index of payment $P$ & Holder of $P$ \\
        $P.\ptime$		& Time when payment $P$ was created & Holder of $P$ \\
        $P.\type$		& Type of payment $P$ (\basic or \conditional) & Holder of $P$ \\
		
		\hline
		\rowcolor{lightgray}
		\textbf{Function} & \textbf{Description} & \textbf{Scope} \\
		\hline
		
		$\H(x)$ 		& Outputs a cryptographic hash of $x$ & Global \\ 
		$\sign(x, \sk)$ & Outputs a signature of $x$ signed with signing key $\sk$ & Global\\
		$\sigverify(x, \sigma, \vk)$ & Outputs $1$ iff signature $\sigma$ over $x$ using verification key $\vk$ is valid & Global\\
		
		$\certverify(\cert, \vk_\S)$ & Outputs $1$ iff $\sigverify(\cert.\vk, \cert.\sig,\vk_\S) \verify 1$ & Global\\
		
		$\hwcertverify(\cert, \vk_\S)$ & Outputs $1$ iff $\sigverify([\cert.\vk, \TA], \cert.\sig,\vk_\S) \verify 1$ & Global\\	

		$\oemcertverify(\vk, \cert, \vk_\D)$ & Outputs $1$ iff $\sigverify([\vk, \text{``Secure Device''}\|\M], \cert,\vk_\D) \verify 1$ & Global\\	

		$\SE_\A.\deposit(x,...)$ & Deposits $x$ amount of money into client $\A$'s secure hardware $\SE_\A$ & $\SE_\A$ \\ 
		$\SE_\A.\withdraw(x,...)$ & Withdraws $x$ amount of money from client $\A$'s secure hardware $\SE_\A$ & $\SE_\A$ \\ 
		$\SE_\A.\pay(x,...)$ & Debits $x$ amount of money from $\SE_\A$ and outputs a payment $P$ & $\SE_\A$ \\ 
		$\SE_\A.\collect(P)$ & Credits a payment $P$ into $\SE_\A$'s balance & $\SE_\A$ \\ 
		
		\hline
%
		\bottomrule
	\end{tabular}
	\caption{Protocol Notations}
	\label{tab:keys}
\end{table*}

\section{The OPS Protocol}\label{sec:opsProtocol}

We now describe the OPS protocol explicitly. We break down the protocol
to its components and explain each thoroughly.

\mahdi{Mention that when a party sends a variable to the other one, the recipient checks the authenticity/validity of the variable and so we do not use temporary variables for simplicity. }


\subsection{Client Setup} \label{sec:setup}
Every client (TEE-enabled or not) needs to participate in a one-time setup protocol to register her device with the server (i.e., establish cryptographic keys and certificates) and to initialize her device's TEE stack in case of a TEE-enabled device.
To register with the server, the client generates a local signing key pair denoted by $(\vk, \sk)$ and submits the verification key to the server. In return, the server initializes the client's account information and returns a certificate denoted by $\cert$ to the client. 

\begin{figure}[H]
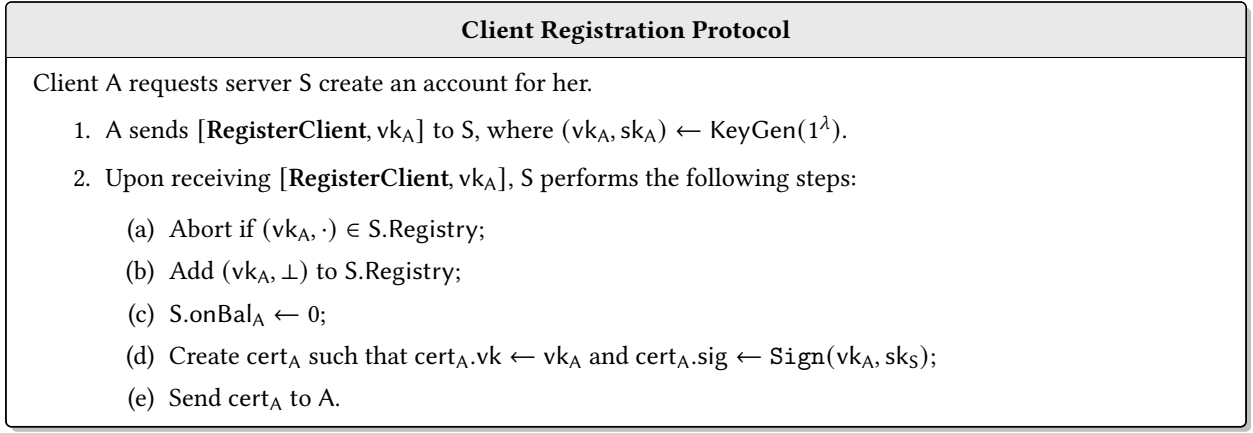

	\begin{shadowbox}[frametitle={Client Registration Protocol}]
	\small			
	Client $\A$ requests server $\S$ create an account for her.
	\begin{enumerate} 			
		\item \A\ sends $[\clientRegisterMsg, \vk_\A]$ to $\S$, where $(\vk_\A, \sk_\A) \gets \keygen(1^\lambda)$.
		\item Upon receiving $[\clientRegisterMsg, \vk_\A]$, $\S$ performs the following steps:			
		\begin{enumerate}
		\item Abort if $(\vk_\A, \cdot) \in \S.\userregistry$;
		\item Add $(\vk_\A, \bot)$ to $\S.\userregistry$;
		\item $\S.\onbal_\A \gets 0;$
		\item Create $\cert_\A$ such that $\cert_\A.\vk \gets \vk_\A$ and $\cert_\A.\sig \gets \sign(\vk_\A, \sk_\S)$;
		\item Send $\cert_\A$ to $\A$.
		\end{enumerate}

	\end{enumerate}
	\end{shadowbox}
	\vspace{-1.5em}
	\caption{Client Registration Protocol}
	\label{fig:clientRegister}
\end{figure}

The certificate is essentially a signature by the server on the client's verification key so that the client can later prove to other entities that her device is registered with the server.
The server maintains a registry (denoted by $\S.\userregistry$) of all registered clients. When a new client registers herself with the server, the verification key of the client is added to this registry. This allows the server to keep track of registered clients in the future to catch duplicate and bogus users. The server also stores the online balance (denoted by $\onbal$) of each registered client. When a new client registers herself with the server, the server initializes an online balance of $0$ for the client. The formal description of this protocol is presented in Figure \ref{fig:clientRegister}.

We now describe how a TEE-enabled client can set up its TEE (according to the GP specification) using remote attestation and TA provisioning.

\paragraph{TEE Remote Attestation.}
The first step to set up a TEE is to obtain an attestation from the original equipment manufacturer (OEM) of the TEE to convince any verifier that the TEE hardware and the trusted OS are authentic.
At a high-level, this is ensured by a read-only memory (ROM), and a device specific \emph{device key-pair} $(\D.\vk, \D.\sk)$.
Both the ROM and the device keys are embedded into the hardware by the OEM.
The trusted OS provides a signature (using $\D.\sk$ from the ROM) on the TEE binaries using a method we denote $TOS.Attest$ along with $\D.\vk$ to the remote party who forwards them to the OEM for verification. Since OEM knows the contents of the TEE stack, it can verify the signature, and hence attest to the authenticity of the TEE using a method we denote $\oemcertverify$.

\paragraph{OPS TA Provisioning.}
Once the TEE is authenticated via remote attestation, the verifier needs to ensure that the OPS TA program (as shown on~\cref{fig:opsTA}) is provisioned (i.e., deployed) properly inside the TEE.
This can be done via either local or remote provisioning~\cite{globalWhitePaper}. In local provisioning, the OEM ships TA binaries on the device as part of the TEE software stack. 
In remote provisioning, a trusted party (e.g., Trustonic~\cite{trustonic:online}) deploys the TA to the TEE remotely after the TEE has been authenticated via the remote attestation process. This is done by first establishing a secure channel with the TEE using the device's verification key, and then transmitting the TA binaries to the TEE over the secure channel. To ensure the TA is deployed properly, the TEE signs a hash of the binary with $\SH.\sk$ and returns the signature to the trusted party for verification.


\subsection{OPS TA Registration}\label{sec:taregister}

After registering with the server, TEE-enabled clients will need to perform two key steps in order to initialize their TEE. First, their TEE must be authenticated by means of (remote) attestation. Next, the OPS TA (as shown in Figure \ref{fig:opsTA}) must be provisioned within their TEE. After the TEE is validated and the OPS TA is setup via provisioning phase,
the client's device and the OPS TA instance need to be registered with the server.
To do this, the OPS TA first generates a signing key pair denoted $(\SE.\vk, \SE.\sk)$, and returns to the client's device the verification key as well as the remote attestation. This process is described by the method $\init$ described in the OPS TA program in Figure \ref{fig:opsTA}. 

Next, the client transmits the TA's verification key and the attestation, along with its own device information to the server. The server, after verifying the attestation (using $\oemcertverify$), certifies the key by signing it with the server's secret key and returning it to the device. 
The signed verification key is a certificate showing that the OPS TA key is generated by a genuine TEE and is registered with the server. Whenever this device makes a payment, it transmits the certificate along with other payment information, so that the receiver can independently verify the validity of the payment using the server's public (verification) key.

The certificate is also required to ``activate'' the client's TEE for offline payments. That is, only after receiving the certificate from the server ascertaining that it has been registered with the server will the OPS TA in the TEE perform any of its functions (other than $\init$). This check is performed by the method $\certinit$ described in the OPS TA program (Figure \ref{fig:opsTA}). Specifically, initializing the variable $\SE.\cert$ by the method $\certinit$ in the OPS TA is necessary for the invocation of other methods.


The server initializes a counter (denoted by $\id$) for each OPS client. The OPS TA also maintains an internal variable denoted by $\SE.\id$. Our protocol ensures that the two counters are ``in sync'' with one another. While the value of the counter would denote the number of deposits or withdrawals that have been performed by the client, the role of this counter is to distinguish various deposits (converting online funds to offline funds) and withdrawals (converting offline funds to online funds) and protect against replay attacks (e.g., replaying a deposit would allow a client to create offline funds out of thin air). This will be explained in further detail in Section \ref{sec:depositwithdraw}.

\begin{figure}[H]
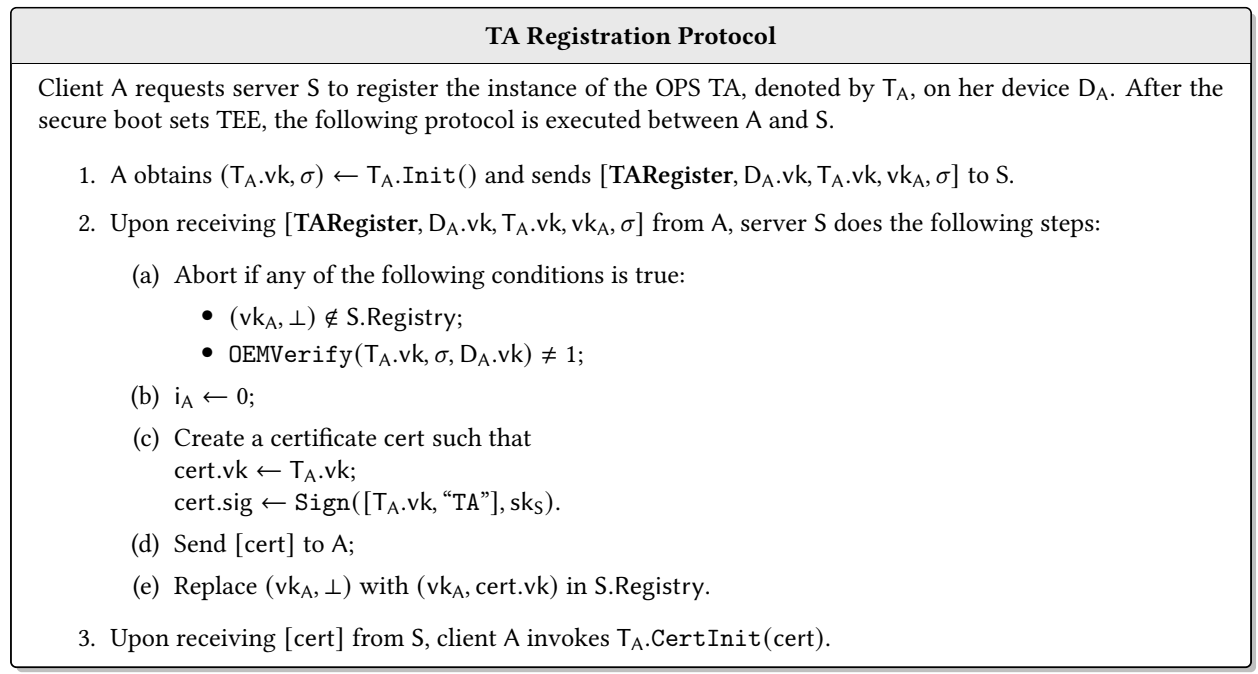

	\begin{shadowbox}[frametitle={TA Registration Protocol}]
		\small			
		Client $\A$ requests server $\S$ to register the instance of the OPS TA, 
		denoted by $\SE_\A$, on her device $\SH_\A$.
		After the secure boot sets TEE, the following protocol is executed between $\A$ and $\S$.
			
		\begin{enumerate} 			
			\item $\A$ obtains	$(\SE_\A.\vk, \sigma) \gets \SE_\A.\init()$
			and sends $[\taRegisterMsg, \SH_\A.\vk, \SE_\A.\vk, \vk_\A, \sigma]$ to S.
					
			\item Upon receiving $[\taRegisterMsg, \SH_\A.\vk, \SE_\A.\vk, \vk_\A, \sigma]$ from $\A$, server $\S$ does the following steps:
			\begin{enumerate}
				\item Abort if any of the following conditions is true:
				\begin{itemize}
					\item $(\vk_\A, \bot) \not\in \S.\userregistry$;  
					\item $\oemcertverify(\SE_\A.\vk, \sigma, \SH_\A.\vk) \ne 1$; 
				\end{itemize}
			
				\item $\id_\A \gets 0$;
				\item Create a certificate $\cert$ such that \\
				$\cert.\vk \gets \SE_\A.\vk$; \\
				$\cert.\sig \gets \sign([\SE_\A.\vk, \text{\TA}], \sk_\S)$.
				\item Send $[\cert]$ to $\A$;
				
				\item Replace $(\vk_\A, \bot)$ with $(\vk_\A, \cert.\vk)$ in $\S.\userregistry$.	
			\end{enumerate}
			
			\item Upon receiving $[\cert]$ from $\S$, client $\A$ 
			invokes $\SE_\A.\certinit(\cert)$.
		\end{enumerate}
	\end{shadowbox}
	\vspace{-1.5em}
	\caption{TA Registration Protocol}
	\label{fig:taRegister}
\end{figure}

The server makes use of its registry to ensure that the client has registered herself before the deposit step. The server also uses the registry to tag the TEE being registered along with the client who is registering it, by storing the pair $(\vk, \SE.\vk)$ in its registry.
The formal description of this protocol is presented in Figure \ref{fig:taRegister}.

\ranjit{We assume that if a client wants to register multiple TAs, then they do so by creating a new UA key pair for each TA key pair. }

\subsection{OPS TA Program}

In Figure~\ref{fig:opsTA}, we present the methods provided by our OPS TA to the TEE-enabled client devices. We first briefly describe them. Note that the workings and the roles of the methods $\init$ and $\certinit$ have already been discussed in Sections \ref{sec:setup} and \ref{sec:taregister}.

\begin{itemize}
	\item $\init$: Initializes the OPS TA, generates a key-pair along with attestation; this is the first method that must be invoked.
	\item $\certinit$: Processes certificate from the server; this is the second method that must be invoked, after which, other methods can be executed.
	\item $\deposit$: Converts online funds into offline funds, increases the offline balance.
	\item $\withdraw$: Converts offline funds into online funds, decreases the offline balance.
	\item $\pay$: Creates an offline payment object.
	\item $\collect$: Verifies an offline payment and applies it to the offline balance by increasing it with the payment amount.
	\item $\getbalance$: Returns the current offline balance stored inside the TEE storage.
\end{itemize}

After registering herself with the server and provisioning the OPS TA on her TEE, the client invokes the $\init$ method of the OPS TA. Using the results from the OPS TA, the client can then register her TEE with $\S$, obtain the certificate from $\S$, and invoke the $\certinit$ method of the OPS TA. Now, the client can:
\begin{itemize}
	\item Convert some/all of her online funds into offline funds in her OPS TA (by invoking the deposit protocol described in Section \ref{sec:depositwithdraw} which will involve invoking the $\deposit$ method of the OPS TA);
	
	\item Use offline funds to make offline payments (by invoking the offline payment protocol described Section \ref{sec:offlinepay} which will involve invoking the $\pay$ method of the OPS TA);
	
	\item Verify offline payments made to her either offline with her OPS TA (by invoking the collect protocol described Section \ref{sec:claimcollect} which will involve invoking the $\collect$ method of the OPS TA) or online with the server (by invoking the claim protocol described Section \ref{sec:claimcollect});
	
	\item Convert some/all of her offline funds into online funds (by invoking the withdraw protocol described Section \ref{sec:depositwithdraw} which will involve invoking the $\withdraw$ method of the OPS TA).
\end{itemize}

\noindent With the overall flow of operations in mind, we now describe the OPS TA program in more detail. The OPS TA maintains some variables whose functions are as follows:

\begin{itemize}
	
	\item $(\SE.\vk, \SE.\sk)$ is the singing key-pair used to authenticate outputs of the OPS TA.
	\item $\SE.\bal$ is the offline balance that is maintained within the OPS TA.
	\item $\SE.\cert$ is the certificate issued by the server on registering the TEE. This certificate allows the OPS TA to be convinced that the TEE has been registered with the server. It is also used by the OPS TA when it generates offline payments to identify itself as an authentic registered sender of offline funds.
	\item $\SE.\iplog$ is the log of offline payments received from other users. It is used to protect a client from a malicious sender who may be replaying a previous payment in an attempt to double-spend.	
	\item $\SE.\id$ is a counter for deposits and withdrawals. The role and workings of this counter were alluded to briefly in Section \ref{fig:taRegister}. Recall that the server also maintains a copy of this counter and the two copies are kept in sync with one another. It is used to prevent replay attacks in the context of deposits and withdrawals. Further details are described in Section \ref{sec:depositwithdraw}.
	\item $\SE.\pid$ is a counter for payments. It is used to make every payment unique. This (in conjunction with the payment log $\iplog$) prevents a client from  replaying a previous payment in an attempt to double-spend.
\end{itemize}

\noindent We now describe in detail the various sub-protocols that are involved in our OPS protocol.

\begin{figure}
	\begin{shadowbox}[frametitle={OPS Trusted Application}]
		\small    
		
		\begin{description}[itemsep=0.75em, leftmargin=2em]
			\item \underline{$\init()$:}
			\vspace{-0.25em}
			\begin{enumerate}
				\item $(\SE.\vk, \SE.\sk) \gets \mathsf{KeyGen}(1^\lambda)$; 
				\enspace $\SE.\bal \gets 0$;
				\enspace $\SE.\cert = \bot;$
				\enspace $\SE.\iplog \gets \bot$; 
				\enspace $\SE.\id \gets 0$; 
				\enspace $\SE.\pid \gets 0$;
				\item $\sigma \gets TOS.Attest(\SE.\vk)$	\mahdi{TODO: What's the actual thing done to get this signature?}
			 	\item Output $(\SE.\vk, \sigma)$.
			\end{enumerate}        	
		        	
			\item \underline{$\certinit(\cert)$:}
			\vspace{-0.25em}
			\begin{enumerate} 
				\item Abort if $\hwcertverify(\cert, \SE.\vk_\S) \ne 1$. 
				\item $\SE.\cert \gets \cert$.
			\end{enumerate}        	
		        	
		    \item \underline{$\deposit(x,\id,\sigma_\S)$:} 
			\vspace{-0.25em}
			\begin{enumerate} 
				\item Abort if $\SE.\cert = \bot$ \orr $\id \neq \SE.\id + 1$ \orr	 $\mathsf{SignVerify}([\SE.\vk, x,\id], \sigma_\S, \vk_\S) \neq 1$;		
				\item $\SE.\bal \gets \SE.\bal + x$;
				\item $\SE.\id \gets \SE.\id + 1$.
			\end{enumerate}
								
			\item \underline{$\withdraw(x)$:} \vspace{-0.5em}
			\begin{enumerate} 
				\item Abort if $\SE.\cert = \bot $ \orr $x > \SE.\bal$;		
				\item $\SE.\bal = \SE.\bal - x$;
				\item $\SE.\id = \SE.\id + 1$;
				\item Output $[x, \SE.\id, \sigma]$, where $\sigma = \sign([x, \SE.\id], \SE.\sk)$. 
			\end{enumerate}
		                        
			\item \underline{$\pay(x,\receiver)$:} 
			\vspace{-0.25em}
			
			\begin{enumerate} 
			    \item Abort if
			    $\SE.\cert = \bot $ \orr  $\SE.\bal < x$;         
			    \item $\SE.\bal \gets \SE.\bal - x$;            
			    \item $\SE.\pid \gets \SE.\pid + 1$;
			                
			    \item 
			    $P.\amount\gets x$; \thinspace
			    $P.\sender \gets \SE.\cert$; \thinspace
			    $P.\receiver \gets \receiver$; \thinspace
			    $P.\index\gets \SE.\pid$;    	
			    \item Output $P$, where $P.\sig \gets \sign([P.\amount, P.\sender, P.\receiver, P.\index], \SE.\sk)$.        
			\end{enumerate}
		
			\item \underline{$\collect(P)$:}
			\vspace{-0.25em}
			\begin{enumerate} 
				\item Abort if $\SE.\cert = \bot $ \orr $\payVerify(P) \neq 1$ \orr $P.\receiver \neq \SE.\cert$ \orr $P \in \SE.\iplog$; 
				\item $\SE.\bal \gets \SE.\bal + P.\amount$;
				\item Append $P$ to $\SE.\iplog$.
			\end{enumerate}	
			
			\item \underline{$\getbalance()$:} 
			\vspace{-0.25em}
			\begin{enumerate}
				\item Abort if $\SE.\cert = \bot$;
				\item Output $[\SE.\bal, \SE.\id, \sigma]$, where $\sigma = \sign([\SE.\bal, \SE.\id], \SE.\sk)$.\\[0.5em]
			\end{enumerate}
			
		\end{description}
	\end{shadowbox}
	\vspace{-1.5em}
	\caption{OPS Trusted Application}
	\label{fig:opsTA}
\end{figure}

\input{deposit-withdraw}

\input{payment-protocol}

\mahdi{Section on Replay Attack Protection -- We use a simple idea of asking the receiver to generate a challenge random value, and then asking the sender to produce a signature on this random value (along with the other payment parameters). This ensures that the sender will need to query the secure element in order to generate the signature, and in particular cannot reuse an older payment signature and double spend. Alternatives?}
%

\mahdi{Section on Payment Confirmation}
%
%

%% file: deposit-withdraw.tex
\subsection{Deposit and Withdraw Protocols} \label{sec:depositwithdraw}
\begin{figure}[H]
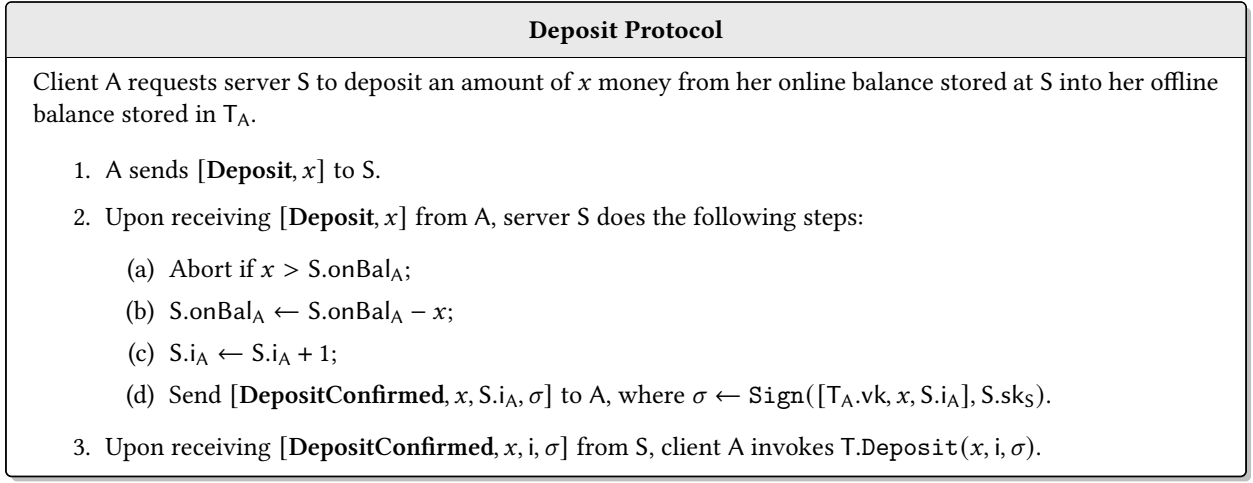

	\begin{shadowbox}[frametitle={Deposit Protocol}]
		\small
		Client $\A$ requests server $\S$ to deposit an amount of $x$ money from her online balance stored at $\S$ into her offline balance stored in $\SE_\A$.
		
		\begin{enumerate}	
			\item $\mathsf{A}$ sends $[\depositMsg, x]$ to $\mathsf{S}$. \mohsen{how does server know this request is from A, there needs to be some $\vk_\A$ or $\SE_\A.\vk,$ along with some signature authorizing this request.}
			\item Upon receiving $[\depositMsg, x]$ from $\A$, server $\S$ does the following steps:
			\begin{enumerate}
				\item Abort if $x > \S.\onbal_\A$;
				\item $\S.\onbal_\A \gets \S.\onbal_\A - x$;
				\item $\S.\id_\A \gets \S.\id_\A + 1$;
				\item Send $[\depositConfirmed, x, \S.\id_\A, \sigma]$ to $\A$, where ${\sigma \gets \sign([\SE_\A.\vk, x, \S.\id_\A], \S.\sk_\S)}$.
			\end{enumerate}
			\item Upon receiving $[\depositConfirmed, x, \id, \sigma]$ from $\S$, client $\A$ invokes $\SE.\deposit(x, \id, \sigma)$.
		\end{enumerate}
		
	\end{shadowbox}
	\vspace{-1.5em}
	\caption{Deposit Protocol (Online $\to$ Offline)}
	\label{fig:deposit}
\end{figure}

\begin{figure}[H]
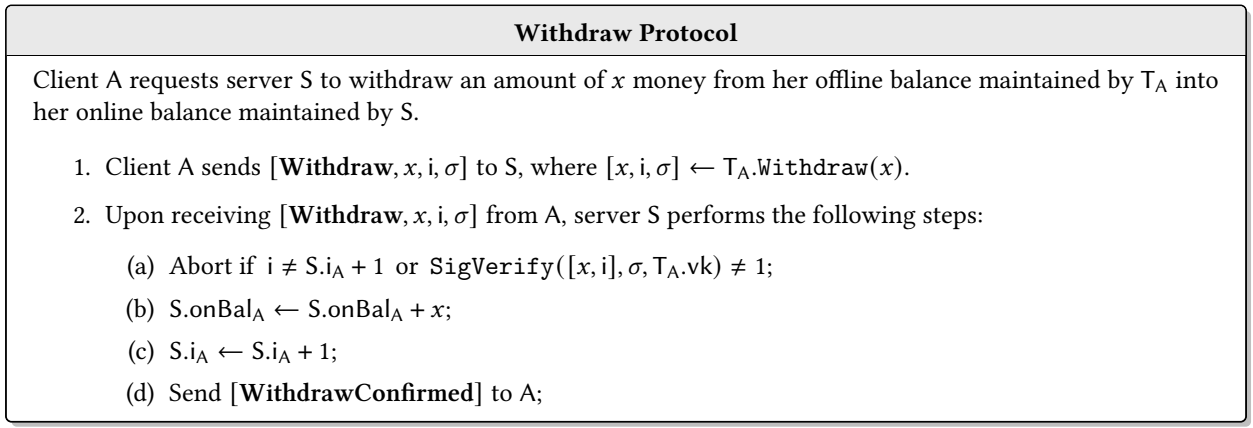

	\begin{shadowbox}[frametitle={Withdraw Protocol}]
		\small		
		Client $\A$ requests server $\S$ to withdraw an amount of $x$ money from her offline balance maintained by $\SE_\A$ into her online balance maintained by $\S$.
		
		\begin{enumerate}			
			\item Client $\A$ sends $[\withdrawMsg, x, \id, \sigma]$ to $\S$, where ${[x, \id, \sigma] \gets \SE_\A.\withdraw(x)}$.
			\mohsen{it seems in all protocols we are simplifying that server knows which message is coming from which client. we have to note it somewhere.}
			
			\item Upon receiving $[\withdrawMsg, x, \id, \sigma]$ from $\A$, server $\S$ performs the following steps:
			\begin{enumerate}
				\item Abort if \medspace $\id \neq \S.\id_\A + 1$ \orr ${\sigverify([x,  \id], \sigma,\SE_\A.\vk) \neq 1}$;
				

				\item $\S.\onbal_\A \gets \S.\onbal_\A + x$;
				
				\item $\S.\id_\A \gets \S.\id_\A + 1$;
				
				\item Send $[\withdrawConfirmed]$ to $\A$;
				
			\end{enumerate}
		\end{enumerate}
	\end{shadowbox}
	\vspace{-1.5em}
	\caption{Withdraw Protocol (Offline $\to$ Online)}
	\label{fig:withdraw}
\end{figure}

In the deposit protocol presented in Figure \ref{fig:deposit}, the client converts some/all of her online funds into offline funds. That is, the client deducts some amount from her online balance, as maintained by the server, and deposits the amount to her offline balance which is maintained by the OPS TA within her device.
The withdraw protocol presented in Figure \ref{fig:withdraw} works in the opposite direction converting offline funds into online funds. That is, the client withdraws some amount from her offline balance and transfers it to the server to add the amount to her online balance. 

The deposit protocol works as follows. The client wishing to deposit an amount $x$ of online funds into her offline balance, sends the request $[\depositMsg, x]$ to the server. The server on identifying the client\footnote{The explicit reference to these details has been omitted in the presentation in Figure \ref{fig:deposit}. This would involve the client identifying themselves for instance using the certificate or verification key which the server could check for in its registry. The server would also need to determine if the client has registered her TEE which would be necessary in order to deposit online funds offline. The client would have to explicitly identify their TEE which the server could then check for in its registry.} checks that the client has sufficient (greater than $x$) online funds. If so, the server deducts an amount of $x$ from the clients online balance and generates a deposit confirmation that contains the amount $x$. Aside from the amount, the confirmation contains two other key pieces of information. The first is the counter $\id$ for deposits and withdrawals, and the second is a signature $\sigma$ by the server on $\SE.\vk$, $x$ and $\id$. We will describe the need for each of these ahead. 
On receiving the deposit confirmation, the client can invoke the $\deposit$ method of the OPS TA with the confirmation. The method checks that the its local copy $\SE.\id$ is ``in sync'' with that of the server (technically, they would be off by $1$ at this stage, but equal to each other once $\deposit$ completes) and that the signature is valid. If so, it increments the offline balance by $x$ and syncs up $\SE.\id$. 

We now describe the role of $\id$ and $\sigma$. The counter $\id$ is used to uniquely identify deposits and withdrawals and thus prevent a client using a particular deposit confirmation more than once. In particular, if a client attempts to invoke the method $\deposit$ using a particular deposit confirmation more than once, the counter $\id$ would be out of sync with that of the server that is present in the confirmation itself. But what if the client attempts to spoof the value of $\id$ in the deposit confirmation in order to make it seem as though it is in sync? This is prevented by the signature $\sigma$. 

Concretely, the presence of $\sigma$ authenticates the confirmation sent by the server. It is not possible for the client to modify the particulars of the confirmation (the TEE, the amount or $\id$) without resulting in an invalid signature in the confirmation. By signing on $\SE.\vk$, deposit confirmations can only be processed by one single (intended) client TEE. By signing on $x$, the TEE is convinced of the authenticity of the deposit amount. By signing on $\id$, as explained before, a deposit confirmation cannot be replayed in an attempt to generate offline funds out of thin air. The formal description of this protocol is presented in Figure \ref{fig:deposit}.

The withdraw protocol works in exactly the same way as the deposit protocol, only in reverse. The client wishing to withdraw an amount $x$ of offline funds into her online balance, invokes the $\withdraw$ method of the OPS TA with the amount $x$. The OPS TA checks that the client has sufficient (greater than $x$) offline funds. If so, the OPS TA deducts an amount of $x$ from the clients offline balance and generates a withdraw confirmation that contains the amount $x$. As in the case of the deposit protocol, the withdraw confirmation also contains the counter $\id$ and a signature $\sigma$ by the OPS TA on $x$ and $\id$. We will describe the need for each of these ahead. 

The client on receiving the withdraw confirmation, sends the request $[\withdrawMsg, x, \id, \sigma]$ to the server. The server on identifying the client checks that the its local copy $\S.\id$ is ``in sync'' with that of the OPS TA and that the signature is valid. If so, it increments the online balance by $x$ and syncs up $\S.\id$. The counter $\id$ prevents a client using a particular withdraw confirmation more than once. The presence of $\sigma$ authenticates the confirmation sent by the OPS TA. It is not possible for the client to modify the particulars of the confirmation (the amount or $\id$) without resulting in an invalid signature in the confirmation. By signing on $x$, the TEE is convinced of the authenticity of the deposit amount. By signing on $\id$, as explained before, a withdraw confirmation cannot be replayed in an attempt to generate online funds out of thin air. The formal description of this protocol is presented in Figure \ref{fig:withdraw}.

%% file: payment-protocol.tex
\subsection{Offline Payment Protocol} \label{sec:offlinepay}

\begin{figure}[H]
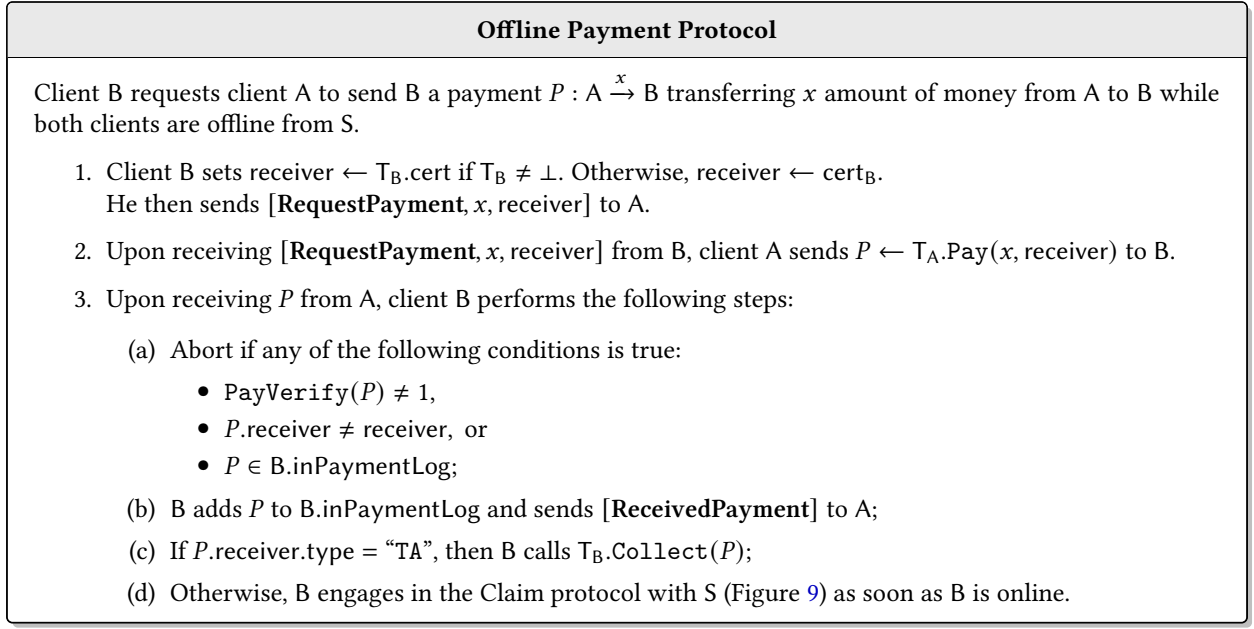

	\begin{shadowbox}[frametitle={Offline Payment Protocol}]
		\small
		
        Client $\B$ requests client $\A$ to send $\B$ a payment ${P: \A \xrightarrow{x} \B}$ transferring $x$ amount of money from $\A$ to $\B$ while both clients are offline from $\S$.		
		\begin{enumerate}     
			\item Client $\B$ sets ${\receiver \gets \SE _\B.\cert}$ if ${\SE_\B \neq \bot}$. Otherwise, ${\receiver \gets \cert_\B}$.\\
			He then sends $[\payMsg, x, \receiver]$ to $\A$.
			
			       
            \item  Upon receiving $[\payMsg, x, \receiver]$ from $\B$, client $\A$ sends $P \gets \SE_\A.\pay(x, \receiver)$ to $\B$.

			\item Upon receiving $P$ from $\A$, client $\B$ performs the following steps:
			\begin{enumerate}
				\item Abort if any of the following conditions is true:		
				\begin{itemize}
					\item ${\payVerify(P) \neq 1}$,
					\item ${P.\receiver \neq \receiver}$, \orr
					\item $P \in \B.\iplog$;
				\end{itemize}
		
				\item $\B$ adds $P$ to $\B.\iplog$ and sends $[\payConfirmed]$ to $\A$;
				
				\item If $P.\receiver.\type = \TA$, then $\B$ calls $\SE_\B.\collect(P)$;
				
				\item Otherwise, $\B$ engages in the Claim protocol with $\S$ (\cref{fig:claim}) as soon as $\B$ is online.
			\end{enumerate}
		\end{enumerate}
	\end{shadowbox}
	\vspace{-1.5em}
	\caption{Offline Payment Protocol}
	\label{fig:pay}
\end{figure}

In the offline payment protocol presented in Figure \ref{fig:pay}, a client makes an offline payment to another client in the following way. First, the receiver of the payment sends a payment request $[\payMsg, x, \receiver]$ to the sender of the payment. The payment request contains the payment amount as well as the certificate of the receiver (denoted by $\receiver$). Upon receiving a payment request, the sender invokes the method $\pay$ of the OPS TA using $x$ and $\receiver$. The OPS TA checks that the sender has sufficient (greater than $x$) offline funds. If so, the OPS TA deducts an amount of $x$ from the clients offline balance and generates a payment confirmation that contains the amount $x$. Aside from the amount $x$, the payment confirmation also contains the certificates of the sender (recall, $\SE.\cert$) and the receiver (recall, $\receiver$), the payment counter $\pid$ and a signature $\sigma$ by the OPS TA on all of these particulars. 

On receiving the payment confirmation, the receiver checks that the payment is valid using $\payVerify$ which verifies the certificates and $\sigma$. In addition, the receiver checks that the payment was intended for her (i.e., matches the receiver) and that it is a fresh payment, i.e., the receiver can maintain a log $\iplog$ of payments she has received, much like the OPS TA does, in order to prevent malicious senders from replaying payments in an attempt to double-spend. If all checks pass, the receiver is convinced of receiving the payment -- at this point, the receiver is ensured to obtain the payment funds. The receiver adds this payment to her log of payments and sends a confirmation message $[\payConfirmed]$ to $\S$. This completes the portion of the protocol that involves the sender making an offline payment. 

In order to obtain the payment funds, the receiver must either invoke the $\collect$ method of the OPS TA, or she is not a TEE-enabled client, she may engage in the claim protocol. Further details regarding this are described in Section \ref{sec:claimcollect}. It is to be noted that offline payments made to TEEs must be collected and not claimed, while payments made to clients must be claimed and not collected. We now describe the need for the various particulars of the payment confirmation. The presence of the sender and receiver certificates in the payment confirmation convince the receiver that the payment is coming from a registered sender and that the payment has indeed been made to herself. The counter $\pid$ prevents a client using a particular payment confirmation more than once. Notice that the $\pay$ method of the OPS TA increments $\pid$. Thus, every payment confirmation generated by a given TEE is unique (at the bare minimum, the value of $\pid$ would differ between them). 

The payment confirmation also contains the sender certificate, this also means that every payment confirmation every generated is unique (either the sender certificate would differ, and if not, $\pid$ would). The presence of $\sigma$ authenticates the confirmation sent by the OPS TA. It is not possible for the client to modify the particulars of the confirmation (the amount, sender and receiver certificates or $\pid$) without resulting in an invalid signature in the confirmation. By signing on $x$ and the certificates of the sender and the receiver, the authenticity of the payment amount and parties involved in the payment is guaranteed. By signing on $\pid$, as explained before, a payment confirmation cannot be replayed in an attempt to double-spend offline funds. The formal description of this protocol is presented in Figure \ref{fig:pay}.

\begin{figure}[H]
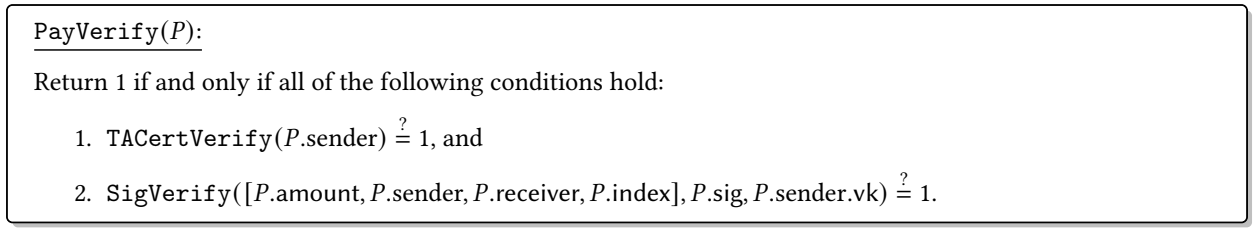

	\begin{shadowbox}
		\small
		\underline{$\payVerify(P)$:} 
				
		\vspace{0.75em}
		Return 1 if and only if all of the following conditions hold:
		\begin{enumerate} 
			\item $\hwcertverify(P.\sender) \verify 1$, and
			\item $\sigverify([P.\amount, P.\sender, P.\receiver, P.\index], P.\sig, P.\sender.\vk) \verify 1$. 				
		\end{enumerate}
	\end{shadowbox}
	\vspace{-1.5em}
	\caption{Payment Verification Method}
	\label{fig:payverify}
\end{figure}


\subsection{Claim and Collect Protocols} \label{sec:claimcollect}

\begin{figure}[H]
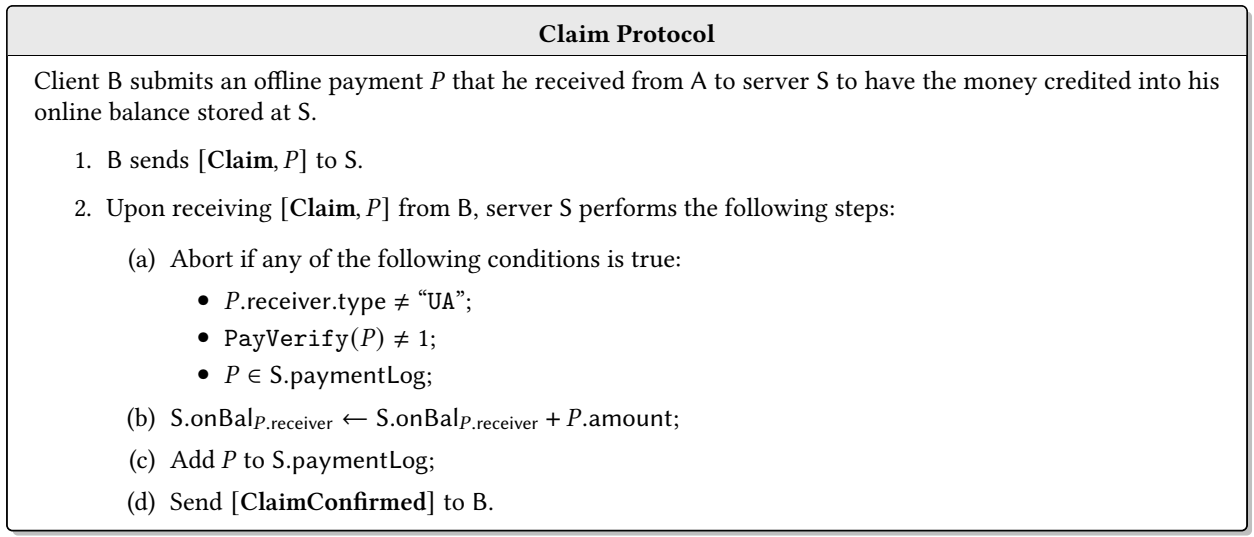

	\begin{shadowbox}[frametitle={Claim Protocol}]
		\small		
		Client $\B$ submits an offline payment $P$ that he received from $\A$ to server $\S$ to have the money credited into his online balance stored at $\S$.
		\begin{enumerate}
			\item $\B$ sends $[\claimMsg, P]$ to $\S$.
			\item Upon receiving $[\claimMsg, P]$ from $\B$, server $\S$ performs the following steps:		
			\begin{enumerate}

				\item Abort if any of the following conditions is true:
				\begin{itemize}
					\item $P.\receiver.\type \neq \UA$; 
					
					
					\item $\payVerify(P) \neq 1$;
					
					\item $P \in \S.\plog$;
				\end{itemize}

				\item $\S.\onbal_{P.\receiver} \gets \S.\onbal_{P.\receiver} + P.\amount$;
				\item Add $P$ to $\S.\plog$;
				\item Send $[\claimConfirmed]$ to $\B$.
			\end{enumerate}
		\end{enumerate}
	\end{shadowbox}
	\vspace{-1.5em}
	\caption{Claim Protocol (Offline $\to$ Online)}
	\label{fig:claim}
\end{figure}

Once a client has obtained a payment confirmation, they can obtain the funds by either invoking the $\collect$ method of the OPS TA, or if she is not a TEE-enabled client, she may engage in the claim protocol presented in Figure \ref{fig:claim} as soon as she is online. As noted in Section \ref{sec:offlinepay}, offline payments made to TEEs must be collected and not claimed, while payments made to clients must be claimed and not collected. This is to prevent a malicious client from collecting and claiming a single payment confirmation in an attempt to generate online or offline funds out of thin air. Thus, a single payment confirmation may only be either collected or claim, but not both. This means that it is sufficient to ensure that a single confirmation cannot be collected more than once, nor claimed more than once.

\paragraph{Collect Protocol.} A client can collect a payment by invoking the $\pay$ method of the OPS TA using the payment confirmation. The OPS TA checks that the payment is valid (performed using $\payVerify$ which verifies the certificates and $\sigma$), that it was intended for itself (matching receiver) and that it is a fresh payment (the OPS TA maintains a log $\iplog$ of payments it has received in order to prevent malicious senders from replaying payments in an attempt to double-spend; this means that a payment cannot be collected more than once). If all checks pass, the OPS TA increments the offline balance by $x$ and adds this payment to its log of payments.

\paragraph{Claim Protocol.} A client can claim a payment by engaging in the claim protocol as shown in~\cref{fig:claim}. The client wishing to claim a payment $P$ sends the request $[\claimMsg, P]$ to $\S$ who checks that the payment is valid and that it is a fresh payment. To achieve this, $\S$ maintains a log $\iplog$ of payments that have been claimed, much like the OPS TA does, in order to prevent malicious clients from replaying payments in an attempt to generate online funds out of thin air. This means that a payment cannot be claimed more than once. If all checks pass, $\S$ increments the online  balance by $x$ and adds this payment to its log of payments. Finally, $\S$ sends a confirmation message $[\claimConfirmed]$ to the client. 

%% file: disclaimers.tex
\section*{Disclaimers}
\itshape
\footnotesize
Case studies, comparisons, statistics, research and recommendations are provided “AS IS” and intended for informational purposes only and should not be relied upon for operational, marketing, legal, technical, tax, financial or other advice.  Visa Inc. neither makes any warranty or representation as to the completeness or accuracy of the information within this document, nor assumes any liability or responsibility that may result from reliance on such information.  The information contained herein is not intended as investment or legal advice, and readers are encouraged to seek the advice of a competent professional where such advice is required.
All trademarks are the property of their respective owners, are used for identification purposes only, and do not necessarily imply product endorsement or affiliation with Visa.